\documentclass[10pt,twocolumn,letterpaper]{article}

\usepackage{cvpr}
\usepackage{times}
\usepackage{epsfig}
\usepackage{graphicx}
\usepackage{amsmath}
\usepackage{amssymb}

\usepackage[breaklinks=true,bookmarks=false]{hyperref}

\cvprfinalcopy % *** Uncomment this line for the final submission

\def\cvprPaperID{****} % *** Enter the CVPR Paper ID here

\begin{document}

%%%%%%%%% TITLE
\title{Selecting Regions of Interest in Large Multi-Scale Images for Cancer Pathology}

\author{Rui Aguiar\\
Stanford University\\
450 Serra Mall, Stanford, CA 94305\\
{\tt\small raguiar2@stanford.edu}
\and
Jon Braatz\\
Stanford University\\
450 Serra Mall, Stanford, CA 94305\\
{\tt\small jfbraatz@stanford.edu}
}

\maketitle
\begin{abstract}
   Recent breakthroughs in object detection and image classification using Convolutional Neural Networks (CNNs) are revolutionizing the state of the art in medical imaging, and microscopy in particular presents abundant opportunities for computer vision algorithms to assist medical professionals in diagnosis of diseases ranging from malaria to cancer. High resolution scans of microscopy slides called Whole Slide Images (WSIs) offer enough information for a cancer pathologist to come to a conclusion regarding cancer presence, subtype, and severity based on measurements of features within the slide image at multiple scales and resolutions. WSIs' extremely high resolutions and feature scales ranging from gross anatomical structures down to cell nuclei preclude the use of standard CNN models for object detection and classification, which have typically been designed for images with dimensions in the hundreds of pixels and with objects on the order of the size of the image itself. We explore parallel approaches based on Reinforcement Learning and Beam Search to learn to progressively zoom into the WSI to detect Regions of Interest (ROIs) in liver pathology slides containing one of two types of liver cancer, namely \textit{Hepatocellular Carcinoma} (HCC) and \textit{Cholangiocarcinoma} (CC). These ROIs can then be presented directly to the pathologist to aid in measurement and diagnosis or be used for automated classification of tumor subtype.
\end{abstract}

%%%%%%%%% BODY TEXT

\section{Introduction}

In the process of cancer diagnosis and classification, a subsection of a tumor is extracted from a patient, given to a pathologist, and scanned and converted into a large, high resolution image called a Whole Slide Image or a digital pathology slide. Because the slide is so large, the pathologist manually searches for and selects patches to observe and aid in the assistance of cancer classification and diagnosis. Often the pathologist will search for patches containing certain anatomical features that will assist them in differentiating between cancer subtypes and making a diagnosis. This process of searching for sets patches according to different objectives works with relatively high accuracy for pathologists with many years of experience, but is often error-prone and may lead to misdiagnosis if more junior pathologists are selecting the patches and using them to make a diagnostic decision. Additionally, the patch selection process itself is time-consuming, even for experienced pathologists. One pathologist with many years of experience we consulted said it took her upwards of twenty minutes to select patches, perform the necessary measurements, and make a diagnosis decision. 

With the problems of patch selection in mind, we set about trying to automate the process to provide a curated selection of patches that would allow  pathologists to make an informed decision about cancer diagnosis, saving them the time and obviating the need for manually navigating across the slide to search for regions of interest. While there has been much work in Artificial Intelligence for cancer classification, from logistic regression to deep learning \cite{sciencedirect1}, we believe that region of interest selection fills a unique niche. We found relatively little in the literature where the AI tool acted as a slide navigation assistant to an experienced pathologist in a routine diagnosis. Instead, many of the papers seemed to be focused on automating the pathologist out entirely through classification and detection of cancer in an image. We seek to go beyond classification and address the problem of how to adapt convolutional neural networks to identify regions of interest in large, multi-scale, whole slide images according to various evaluation metrics that pathologists use when navigating digital pathology slides. To begin, we started with slides containing samples of two types of liver cancer: \textit{Hepatocellular Carcinoma} (HCC) and \textit{Cholangiocarcinoma} (CC). We used a DQN (Deep Q Network) to help automate this patch selection process, in parallel with a CNN + Beam search approach. The input to our model is a whole-slide image (WSI), our network is a DQN/CNN with beam search, and the output is a set of $k$ patches that our model believes will inform a cancer diagnosis.
%-------------------------------------------------------------------------
\section{Related Work/Literature Review}
As mentioned in the introduction, cancer subtyping and detection is a popular application of Artificial Intelligence and computer vision. The classification task for HCC using computer vision has been addressed in \cite{Atupelage1}. This paper used manual feature annotation on the types of nuclei in a cell. From there, they segment the nuclei out and extract four features - inner texture, geometry, spatial distribution, and surrounding texture. From there, the paper uses a SVM on these features to perform cancer classification.

Segmentation across whole-slide images has been used to assist pathologists with cancer diagnosis, as seen in Zhizhao et al \cite{Zizhao}. Here, the authors created a piece of software, where, using RNN's and CNN's, allowed pathologist to enter features they were interested in seeing inside of a whole-slide image, and returned patches that had those relevant features. For example, a pathologist could enter "malformed nuclei" and the software would return patches of the slide where it believed there were malformed nuclei that may correspond to cancer. This paper also performed a whole-slide segmentation, an expensive operation that attempted to highlight areas of the entire slide that had a high probability of being cancerous. This is a clever approach that approaches the state of the art in terms of interpretability to pathologists, though it focuses more on segmentation than returning a series of interpretable patches.

Reinforcement learning has been used to select positions on an image where cancer is suspected to be before as well. One paper that addresses this topic is by Maicas et al \cite{Maicas}. This paper uses a DQN\cite{dqn} on an image that contains a section of breast cancer and attempts to draw a rectangular box around the area it believes the cancer to be present. The actions for this DQN are moving the bounding box around the image, scaling the image and selecting a patch. The reward the network receives is the dice coefficient in the region it selects - that is, the number of pixels that contain cancer divided by the number of pixels in the bounding box. This paper is a similar task and methodology, however, it is not directly applicable to whole-slide images which most pathologists use regularly, and does not address navigating zoom levels thoroughly. 

\cite{tool} analyzed WSIs using CNNs for the tasks of prostate cancer identification in biopsy specimens and breast cancer metastasis detection in sentinel lymph nodes to improve the efficacy of prostate cancer diagnosis and breast cancer staging. They found that, by using a 10-layer CNN, they could identify all slides containing prostate cancer and micro- and macro-metastases of breast cancer could be identified automatically while 30-40\% of the slides containing benign and normal tissue could be excluded without the use of any additional immunohistochemical markers or human intervention. 

\cite{mitosis} was the first to apply convolutional neural
networks to the task of mitosis counting for primary breast cancer grading. Furthermore, in a different publication, they showed the applicability of patch-driven convolutional neural networks to segmentation tasks. Their networks were trained to classify each pixel in the images, using as context a patch centered on the pixel. Simple postprocessing was then applied to the network output.

\cite{glioma} proposed a deep learning-based, modular classification pipeline for automated grading of brain gliomas, the most common class of primary malignant brain tumors using digital pathology images. Whole tissue digitized images of pathology slides obtained from The Cancer Genome Atlas (TCGA, \cite{tcga}), an open source cancer dataset, were used to train their deep learning modules. Convolutional Neural Networks are trained for each module for each sub-task with more than 90\% classification accuracies on validation data set, and achieved classification accuracy of 96\% for the task of GBM vs LGG classification, 71\% for further identifying the grade of LGG into Grade II or Grade III on independent data set coming from new patients from the multi-institutional repository.

\cite{detect} presents a reinforcement learning approach for detecting objects within an image. Similar to our approach of selecting tiles in a WSI to zoom into, their approach
performs a step-wise deformation of a bounding box with the goal of tightly framing the object to be detected, and like our approach it
also uses a hierarchical tree-like representation of predefined region candidates, which the agent can zoom
in on. Their best performing approach comprises a zoom stage and a bounding-box refinement stage, uses aspect-ratio modifying actions and is trained using a combination of three different reward metrics. Our reward metrics are simpler than the ones used in this paper, and we also constrain our RL agent to only zoom actions rather as opposed to also including the option to refine the bounding box.

For our technical approach, we drew inspiration from Deep-Q Networks \cite{dqn} and ResNet \cite{resnet}. The Deep-Q network is a network that takes in an image and learns a Q function from a given reward structure that determines which actions an agent should take. ResNet is a commonly used and cited neural network that includes residual connections to address the issue of a vanishing gradient.
%-------------------------------------------------------------------------
\section{Methods}
%Define your problem precisely, including inputs and outputs of your method.

Our method takes as input a liver cancer pathology whole slide image with dimensions 116143x76502 in the .svs format and outputs a set of small cropped patches of size YxY that contain high amounts of cancerous cells. The image is broken up into 64x64 pixel tiles representing candidate regions of interest according to a task-specific metric, and the output is some number $k$ of tiles that should rank among the highest out of all tiles in the image according to that metric. For concreteness in this report, we will focus on the task of efficiently choosing patches of the image that contain cancerous cells.

Normally the task of finding regions containing cancerous cells would be a standard image segmentation problem. The size of our WSIs, however, make standard image segmentation approaches intractably expensive to perform, so we settle for the weaker task of finding some number $k$ of regions containing a high percentage of cancer cells. In order to cope with the large size of the WSIs, we used an approach similar to \cite{detect} and \cite{glioma} and make use of the hierarchical nature of high-resolution images to iteratively zoom in on promising areas of the slide. For any region of the slide at some magnification that we are considering, we will decide in which subregion to zoom further by evaluating a ``score'' of each subregion and choosing to zoom in on the one with the highest score. Our score is given by our reward function, which we discuss in the Deep-Q approach section. We restrict our candidate subregions to be the four quadrants of the region being considererd rather than choosing arbitrary bounding boxes as done in \cite{Maicas}. Depending on how we choose this ``scoring function'', this can be either considered as a reinforcement learning problem or a tree search problem. The scoring function in either case is a neural network that learns what a good score function should be to maximize the score of the final tile that the agent chooses at the highest zoom level, and the distinction between viewing this as an RL problem or a tree search problem is the same as the distinction between learning the scoring function as the agent chooses zoom trajectories and learning the scoring function beforehand in a supervised-learning fashion.

To be clear, there are two kinds of models that we are considering and evaluating. We have a model that has as input a WSI and outputs a set of tiles at the highest magnification level that should maximize some ground-truth value associated with each one (percentage of the image that are cancerous cells for our task as found from segmentation maps), and we have a submodel that acts as the ``scoring function'' guiding where the agent is zooming to within the slide. The evaluation metric for the primary task that we are trying to maximize is the average ground-truth score over all final selected patches, and the evaluation metric for the scoring model is the MSE of the output with respect to the ground-truth scores.

To answer this question, we clearly defined our input as: a whole-slide image in .svs format, which is a common file format that most digital pathology scanners support. We then broke down that .svs image into many different patches to feed to our score function, which estimates the probability of a patch being indicative of cancer and useful to a pathologist for diagnosis. With our new dataset of patches to a score function, we trained a baseline regressor, a DQN and a ResNet \cite{resnet}. We our loss function on these ground-truth patch scores to estimate a score given a new patch. We then fed that network downsampled quadrants of the whole slide images and choose the quadrant with the highest score to zoom in on, effectively doing a recursive tree search to choose the best patches and output a set $P = \{x^1, x^2...\} \in R^m$ where each element in P is a patch that mostly contains cancer cells as opposed to normal cells or slide background.

\begin{figure}[t]
\begin{center}
   \includegraphics[width=0.9\linewidth]{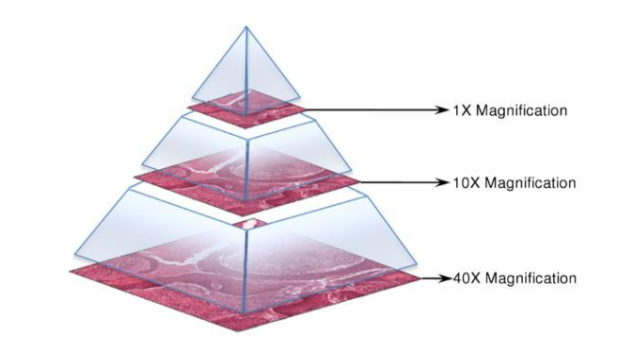}
\end{center}
   \caption{Hierarchical structure of a Whole Slide Image}
\label{fig:long}
\label{fig:onecol}
\end{figure}

\subsection{Methods - Technical Approach}

Note: For all of the approaches, we normalize the images before inputting them into the network with the following equation $$x := (x-\mu_x)/(\sigma)$$ where $\mu_x, \sigma$ are the imagenet mean and standard deviation from pytorch \cite{Pytorch}.

\subsubsection{Baseline Technical Approach}
For our baseline, we used a linear regressor that trained with stochastic gradient descent using squared loss $$\text{Loss} = \frac{1}{n}\sum_{i=1}^n(Y_i-\hat{Y}_i)^2$$
Additionally, we introduced a L2 regularization term $$\text{reg} = \lambda\sum_{j=1}^p\beta_j^2$$ to reduce overfitting. Our algorithm took in as input a patch image represented as a numpy array and tried to output a percent cancer ranging from 0-1. We used SGD on this model to train it, with a constant learning rate of $\alpha = 10^{-5}$. After training this model, we ran the regressor over every patch in a test slide and returned the top 10\%. We then took the average percentage cancer of these top patches and used that as our primary evaluation metric.

\subsubsection{Deep-Q Technical Approach}
We model the Region of Interest search problem formally as a deterministic Markov Decision Process defined by the 4-tuple $(S, A, T, R)$, where
\begin{itemize}
    \item $S$ is a finite set of states representing DeepZoom tiles for a particular WSI.
    \item $A$ is a finite set of actions corresponding to zooming into each of the 4 quadrants of the current tile.
    \item $T$ is a transition function mapping state $s$ and action $a$ to the next state $s' = T(s, a)$ after taking action $a$ at state $s$.
    \item $R$ is a reward function mapping the state $s$ to the immediate reward $R(s)$ upon transitioning to that state.
\end{itemize}

The metric for the "interestingness" of a particular region is given by the reward function $R$. For our experiments we chose the task of finding regions containing cancerous cells, so $R(s)$ for a tile, or a 64x64 patch of the slide $s$ is the fraction of the image containing cancerous cells. That is, the "score" can be mathamatically defined as $$R(s) = \frac{\text{Pixels with cancer present}}{\text{total pixels of area}}$$ This metric optimizes selecting patches with a high percentage of cancer, and we believe this metric is most useful for the pathology-assistant context. As noted in \cite{dqn} we attempted to minimize the following loss function at every iteration $i$: $i$ is $$L_i(\theta_i) = E_{s, a~p(.)}[(y_i-Q(s, a;\theta_i))^2]$$ Where $\theta_i$ are the weights from the current iteration, $y_i$ is the target for iteration $i$, s is the state and a is the action taken. We also introduce epsilon-greedy behavior for our DQN, where it performs the following algorithm at each step \[
  \text{With probability}\begin{cases}
               \epsilon \text{ Take a random action}\\
               1-\epsilon \text{ Take the optimal action}
            \end{cases}
\]

This allowed us to explore random paths and construct an accurate representation of rewards

Our DQN took the form of a small CNN structured as $$[input -> conv -> conv -> fc]$$ to learn the Q function $Q(s, a)$, and then used this CNN at test time on new WSIs to determine our policy, similar to DeepMind's Deep Q-Network for playing Atari games. \cite{dqn}. We experimented with several different networks, but found that this structure achieved the highest rewards on our test set.
\subsubsection{CNN Technical Approach}

As an alternative to learning a Q-function with which to determine a zoom policy via epsilon-greedy exploration, we also implemented a regression-based supervised learning model that directly maps tiles to the percentage of that tile that is cancerous using a ResNet. The input tiles were selected randomly from the same set of training slides that the DQN was trained on, and the labels were the same ground-truth percent-cancer labels that we used to generate the reward signal for the DQN. At test-time, the 4 quadrants of the current tile are passed to the ResNet and we choose to zoom in on the quadrant that is the argmax of the network's output, continuing in this fashion until we've obtained a tile at the highest magnification. To select many high-scoring tiles, we use beam search, which also mitigates the risk of choosing a low-scoring quadrant early in the trajectory that would doom the agent to achieve a low score.

The difference between this supervised method and the reinforcement learning based method is the reward structure and the data that is used to train each method's CNN. The value for a particular tile for the DQN is equal to the percentage of the tile that is cancerous plus the maximum of the values for that tile's four quadrants as formalized in the Bellman equation. Namely, at each step the DQN is evaluating a Q-function $Q(s, a)$ for tile $s$ and zoom-action $a$ representing zooming to one of the 4 quadrants equal to 
\begin{equation*}
    Q(s, a) = R(s') + max_a' Q(s', a'),
\end{equation*}
where $s'$ is the tile in the quadrant associated action $a$. On the other hand, the values for tiles using the supervised method is simply the percentage of that tile that is cancerous, so  
\begin{equation*}
    Q(s, a) = R(s')
\end{equation*} in the terminology we've used for the DQN. In other words, the CNN that the DQN learns the discounted sum of rewards until the end of the trajectory for each action, whereas the supervised learning method learns the rewards directly.

If the Q-network rewards were 0 except at the highest magnification tiles at which point they were equal to the percentage of the slide that is cancerous, then the value function for a tile at a low magnification would be equal to the maximum score of a leaf tile within that region, which is effectively the same as a binary 0/1 loss for whether that region contains any cancer at all, or 
\[
  \text{Loss=}\begin{cases}
               1 \text{ No cancer present}\\
               0 \text{ Cancer present}
            \end{cases}
\]. 

However, we did not use this equation. However, we did not use this loss equation, as it is not differentiable. Instead, we replaced the max function over the leaf scores with an average, giving us the value function that the CNN model is learning.

In summary, we have a tree with leaf nodes with certain scores, and we want to train a CNN to guide the tree search in such a way that we can find the leaf nodes with the highest scores without actually looking at most of the leaf nodes.

\begin{figure}[t]
\begin{center}
   \includegraphics[width=0.6\linewidth]{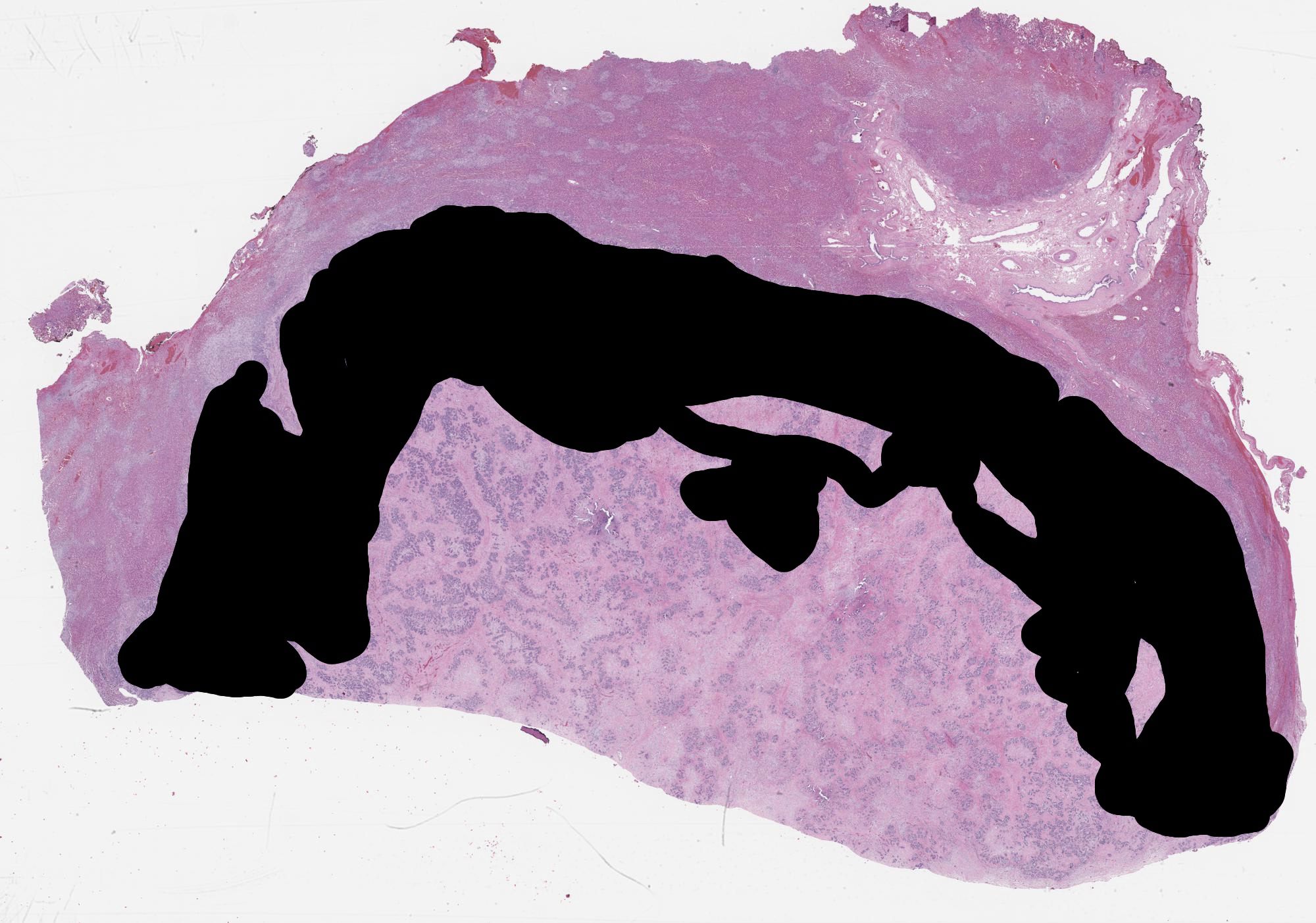}
\end{center}
   \caption{Whole-slide image of cancer, with a segmentation map overlaid on top}
\label{fig:long}
\label{fig:onecol}
\end{figure}

\section{Dataset and features}

Our dataset consists of 49 hematoxylin and eosin (H\&E) stained whole slide images from the The Cancer Genome Atlas (TCGA) publicly available hepatocellular carcinoma (LIHC) and cholangiocarcinoma (CHOL) diagnostic FFPE WSI collections \cite{tcga}. Each svs file contains image scans captured at 4 different magnifications of increasing magnitude. We used 35 of the slides of our training set, 2 for our validation set, and 2 for our test set for our CNN. For our DQN, we used We used 39 of the slides of our training set and 10 for our test set. We use the OpenSlide Python library for interacting with Whole Slide Images, and specifically its DeepZoom functionality for addressing tiles by its level of magnification and position within the image. While images of each slide were captured at four different magnifications, DeepZoom provided the ability to tile a slide image in a grid of size equal to a power of two and increase magnifications by arbitrary powers of two. In particular, from any tile at a particular magnification, we can easily obtain tiles corresponding to each quadrant at twice the magnification, which correspond to the actions in our MDP. The candidate regions of interest from our datset are patches from these WSIs at 10x magnification and with sizes of 64 x 64 pixels.

\begin{figure}[t]
\begin{center}
   \includegraphics[width=0.6\linewidth]{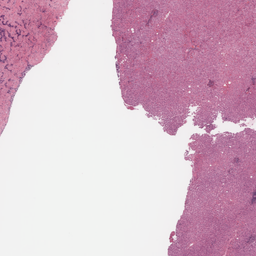}
\end{center}
   \caption{An example patch from a whole-slide image}
\label{fig:long}
\label{fig:onecol}
\end{figure}

\subsection{Preprocessing Pipeline}
We've obtained segmentation maps for each WSI showing which regions of the slides contain cancerous cells. These segmentation maps were created manually by pathologists from the Stanford medical school and are in the form of .jpg images of the downsampled slide image, with cancerous regions being colored black (i.e pixel values of 0 or 1). We calculated scores for each tile at each magnification level by iterating through the pixels in each tile and calculating the fraction of pixels in each tile that were black. This is our reward function $R(s)$ for the each tile $s$. We then stored the tile coordinates and the reward in a CSV to be read in by our RL environment to provide rewards for the DQN.

Our dataset initially consisted of 49 svs slides. From here, we perform prepossessing across each slide to generate a set of patches, the coordinates and zoom level for each patch, and an associated score for each as well. Now, we have a set of data points $D_{train} = \{(x^1, y^1), (x^2, y^2), ...\}$ where $y^i$ is the scorer for datapoint i, and $x^i$ is the jpg image of a patch we extracted from the whole-slide image. Additionally, we perform normalization on these patches that we use as input to our networks. We end up with approximately 49000 patches from these slides with associated scores we train our model on.

\section{Experiments/Results/Discussion}

\begin{figure}[t]
\begin{center}
   \includegraphics[width=0.9\linewidth]{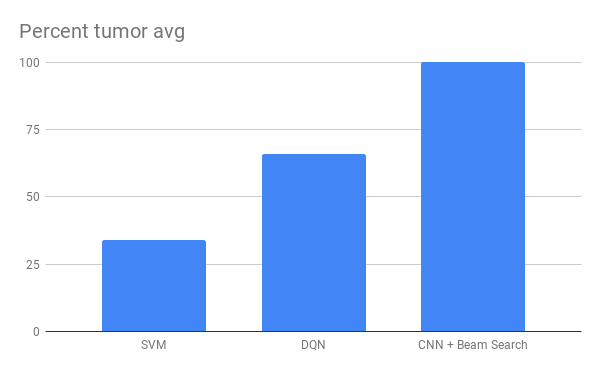}
\end{center}
   \caption{Different model performance comparison - that is, the average percent cancer on patches returned by the model }
\label{fig:long}
\label{fig:onecol}
\end{figure}

The baseline method that we compare our approach with takes as input a patch and outputs the predicted percent cancer in that patch. If we apply this to every patch in a whole-slide image and take the top 5-10\% of patches, some fraction of each selected patch will be cancerous cells, and our evaluation metric is this average fraction among all the patches selected by the model. For example, if our evaluation metric is .34, that means that on average, our baseline was returning patches that contained 34\% cancer.

Our baseline model is a simple linear regressor that takes as input a patch and outputs the predicted percent of cancer in that patch. It was trained with SGD, and outputs, on average, 34\% cancer. This is better than random, which was outputing patches with 20\% cancer, but there is still room for signifigant improvement. 

After implementing our baseline, we experimented with two separate models: A Deep-Q network, and a CNN + beam search combo. For the DQN, the metrics we looked at were the aggregate reward generated over zooming in on a slide, and the average percent cancer it returned from looking at the patches in our test set. For the CNN + beam search combo, we also looked at the average percent cancer it returned from looking at the patches in our test set.

\subsection{Deep-Q}

\begin{figure}[t]
\begin{center}
   \includegraphics[width=0.9\linewidth]{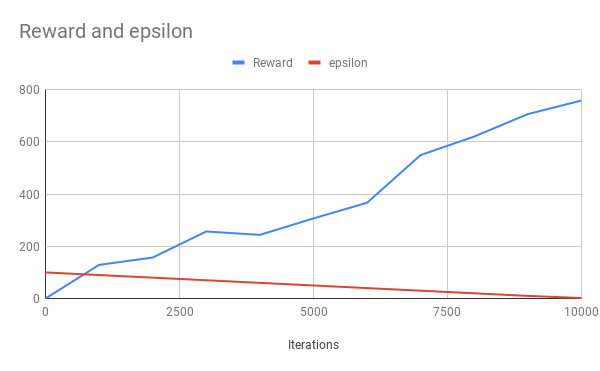}
\end{center}
   \caption{Deep-Q Reward Progression}
\label{fig:long}
\label{fig:onecol}
\end{figure}
As our first approach, we decided to model the problem using reinforcement learning, and to use a Deep-Q network to determine the optimal patches to select given an image of a slide. As you can see from Figure 4, the Deep-Q network performed substantially better than the regressor - that is, it was able to return patches with a higher percentage of cancer contained inside of them. Instead of using a loss curve to evaluate our DQN, we measured performance by looking at how the reward increased over time. This graph is displayed in figure 4. As you can see, the reward increases with the number of iterations as our exploration percentage decreases, showing we are converging to some kind of optimal solution. Through training, we increased our reward from 0 to 759 out of a theoretical maximum of 900. Also displayed in this graph is epsilon - that is, the time spent exploring vs exploiting. You can see that as we explored less our reward increased, indicating we were taking high-reward paths over time. One other thing to note here is that we actually used several different networks for our DQN, including ResNet. The differences in total reward is illustrated in figure 6. We found that using anything larger than our simple small convolutional neural network took too long to converge, and often produced suboptimal results.

\subsection{CNN + Beam search}

Our supervised CNN that learns the percentage of a tile image that is cancerous was a ResNet-50 model, trained over 35 WSIs with a batch size of 64 randomly chosen tiles from the image. We were able to get an accuracy of greater than 99\% on a validation set of consisting of randomly chosen tiles from two slides not in the training set. This greatly outperformed our baseline, and also outperformed our DQN. The ResNet was pretrained using ImageNet weights. At test time, we selected two other withheld slides and ran beam search on the tree of zoom trajectories, with values at each node given by the output of the ResNet. We chose $k=1000$ and compared the number of leaf-level patches selected with their ground-truth percent-cancer labels, and found that the average percentage of the chosen slides that were actually cancer was fully 100\%. A histogram of the outputs on the validation set is given by figure 11, and a sample patch returned is illustrated in figure 7.

\subsection{Hyperparameters}
For our baseline, we chose a stopping criterion of $10^{-3}$, a learning rate of 0.0001 and L2 regularization on the linear regressor trained with SGD. For our DQN, we used a learning rate of $5*10^{-4}$, a batch size of 32, 100000 iterations, and no prioritized replay. We began with 90\% exploration and ended with 2\% exploration across our 100000 iterations. We used a discount factor of $\gamma = 1.0$.

The CNN was trained using the default ResNet-50 from the Keras.applications package using Adam optimization, namely a learning rate of .001, $\beta_1 = .9$, $\beta_2 = .999$, $\epsilon = 0$, and $decay = 0$. It was trained over 10 slides with 5000 randomly sampled tiles per slide at each magnification level for 9 epochs.

\subsection{Experiments}
\begin{figure}[t]
\begin{center}
   \includegraphics[width=0.9\linewidth]{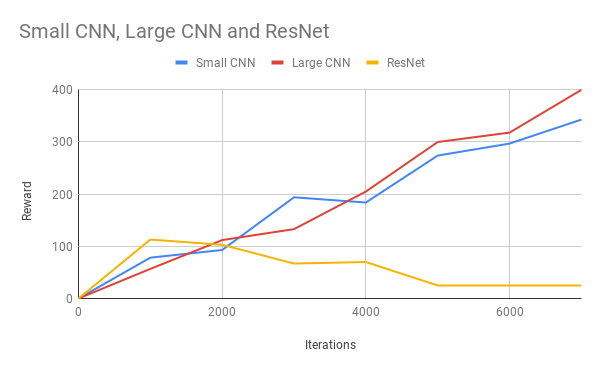}
\end{center}
   \caption{The rewards different networks receive during one training session of our DQN agent}
\label{fig:long}
\label{fig:onecol}
\end{figure}

DQN experiments: For each of our DQN experiments, we input 10 slides and output 10 patches - one from each slide. Our evaluation metric is the percent cancer returned in the final patch, as well as the sum of rewards as it made its decisions to zoom in and select. We experimented with several parameters here, but the one that most drastically affected our performance was varying the network that we used to learn our Q value. The results are illustrated in figure (5). There are a couple of interesting things to notice here. First, the small CNN (2 convolutional layers and 1 FC layer) and large CNN (4 convolutional and 2 FC layers) actually receive similar rewards during their training. However, ResNet converges to a suboptimal solution. This may be due to several reasons, such as not having enough exploration time or iterations to have the network adjust to any kind of reasonable baseline before converging on a suboptimal solution.

One other experiment we ran was using prioritized replay for our DQN, as specified in \cite{prireplay}. Prioritized replay is a form of experience replay that attempts to prioritize more important experiences to replay to the network while training. However, we did not see a material difference when using prioritized replay versus not using experience replay at all.  

CNN experiment: For our CNN, we experimented with different networks. We tried using a small CNN with 2 convolutional layers and 1 fully connected layer, as well as ResNet. We found that ResNet had far higher performance.

\subsection{Results}
Figure 8 illustrates a sample of our results when we had the RL agent and DQN make patch selection decisions about the cancer present in slides it had never seen before. You can see that, as illustrated by the last\_reward column, the agent performs well for many of the slides in the test set. Additionally, the sum of the rewards is relatively high for these slides, indicating that the agent is actually selecting patches with a high cancer percentage at each level, meaning that it is returning interpretable patches at multiple zoom levels when it does detect cancer in the cell.

\subsubsection{CNN Results}
We found that a ResNet-directed beam search of a test slide with $k=100$ generated 100 patches that were entirely within the canerous region of the slide, and when setting $k=1000$ to select 1000 cancerous tiles, we found that the average percentage of each tile that was cancerous was 98.4\%, with 15 of the selected tiles containing no cancerous cells, 1 having 97\% cancerous cells, and, 984 being completely cancerous. A random sample of 1000 patches at the lowest magnification yields a distribution of 740 patches with no cancerous cells, 255 patches with entirely cancerous cells, and 5 patches with intermediate values. This shows that our CNN-directed tree-search method is an effective way to learn a segmentation model and that reinforcement learning shows no advantage over supervised learning trained with random selections of patches at all magnifications. This result held true regardless of whether we normalized the images pixel wise to have a mean of 0 and standard deviation of 1.

\begin{figure}[t]
\begin{center}
   \includegraphics[width=0.4\linewidth]{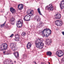}
\end{center}
   \caption{A patch our model correctly returns as 100\% cancer}
\label{fig:long}
\label{fig:onecol}
\end{figure}

\begin{figure}[t]
\begin{center}
   \includegraphics[width=1.0\linewidth]{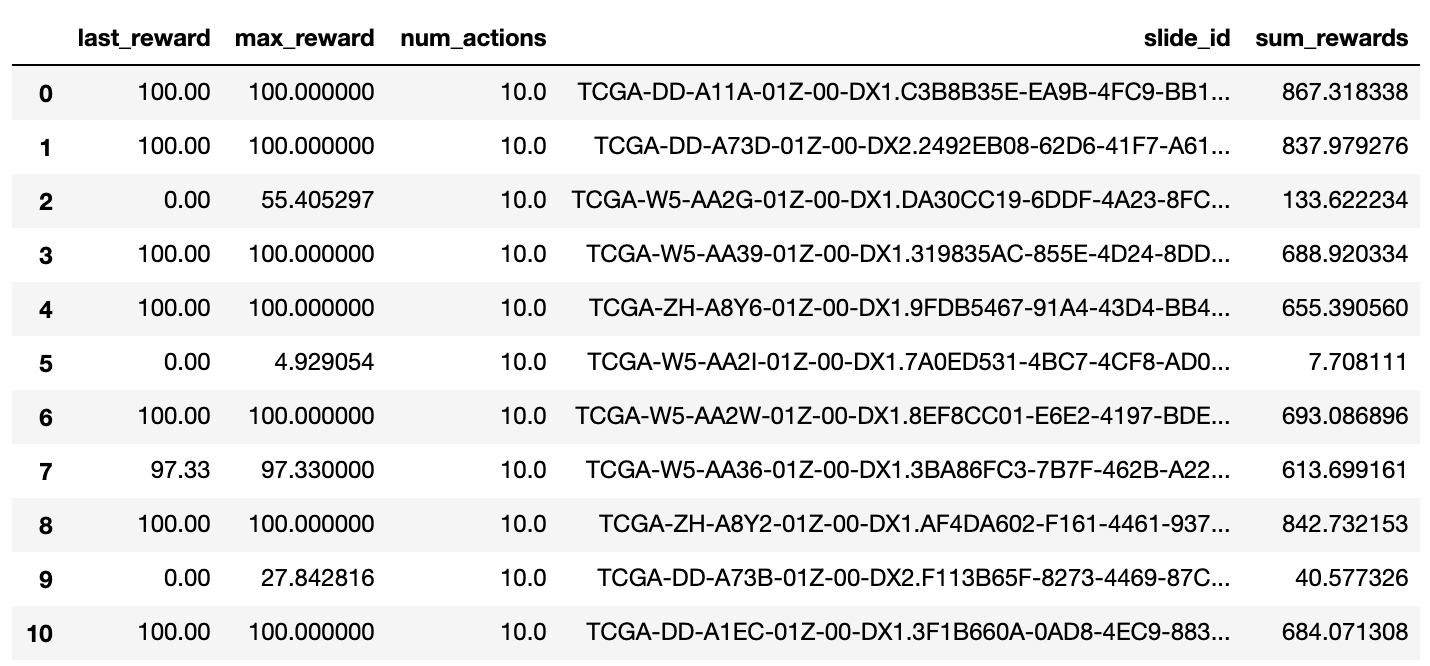}
\end{center}
   \caption{A sample of rewards and patches generated by our model}
\label{fig:long}
\label{fig:onecol}
\end{figure}

\begin{figure}[t]
\begin{center}
   \includegraphics[width=0.9\linewidth]{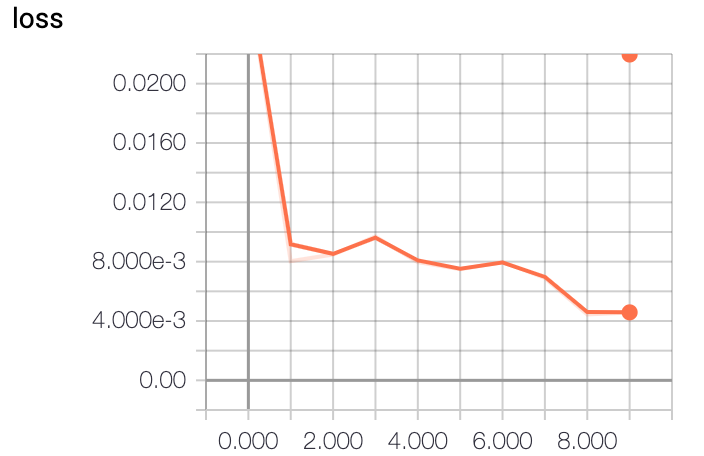}
\end{center}
   \caption{Training loss curve from our beam search}
\label{fig:long}
\label{fig:onecol}
\end{figure}

\begin{figure}[t]
\begin{center}
   \includegraphics[width=0.9\linewidth]{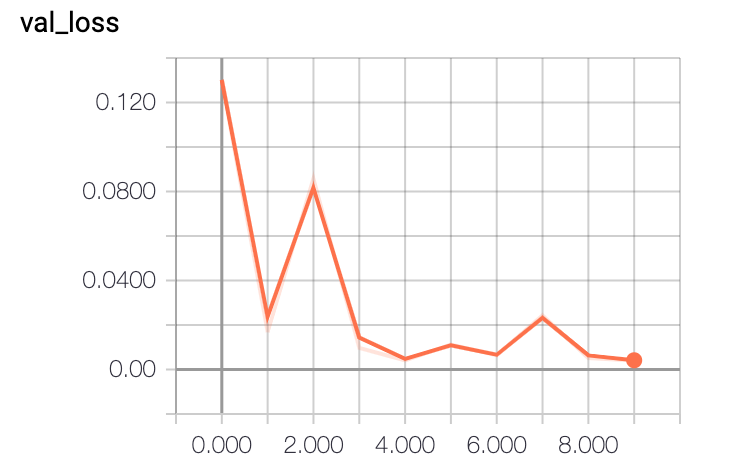}
\end{center}
   \caption{Validation loss curve from our beam search}
\label{fig:long}
\label{fig:onecol}
\end{figure}

\begin{figure}[t]
\begin{center}
   \includegraphics[width=0.9\linewidth]{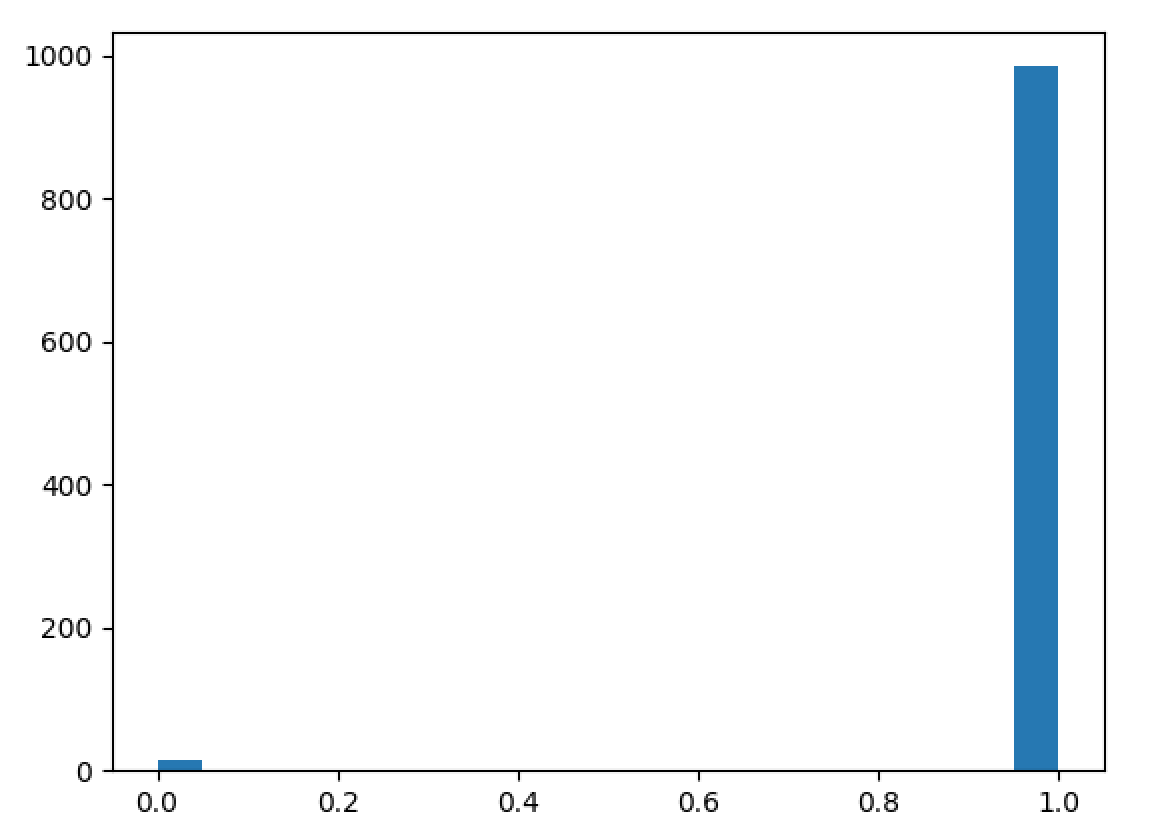}
\end{center}
   \caption{Histogram of accuracy from our beam search}
\label{fig:long}
\label{fig:onecol}
\end{figure}

\subsection{Error Analysis}

There is one important caveat to note in the results for the DQN, and that is that there are some slides where we not only return patches without cancer, but also have a low reward at each step of the process. This happens more specifically at examples (2), (5) and (9). We decided to investigate more into these specific slides to understand why our agent was making sub-optimal decisions. We found that for example (2), the slide was of type CC, which was only a small subset of our training examples. It's very possible that the classifier failed to generalize well to CC slides. Examples (5) and (9) were both HCC, but when we were observing the cancer detection agent at work, we noticed that the agent tended to zoom into sections that were similar in color to cancerous regions, but happened to be healthy. This indicates that our model is learning some of the features associated with cancer, such as a darker pink color on the slide or larger nuclei, but is sometimes confusing healthy cells that may look slightly cancerous with true cancerous ones. 

%Summarize your report and reiterate key points. Which algorithms were the highest-performing?  Why do you think that some algorithms worked better than others? For future work, if you had more time, more team members, or more compute, what would you explore?
\section{Conclusion/Future Work}
Out of our three algorithms, the regressor baseline was the least effective, the DQN the second-most effective, and the CNN with beam search was the most effective. The regressor was likely too simplistic to learn the complicated features needed to make an accurate prediction on cancer diagnosis, and therefore performed poorly. The DQN performed slightly better, but still underperformed the CNN with beam search. Possible reasons for this may include not enough time training the agent, or perhaps a rewards structure that was did not encompass the problem accurately. With more time, we could have attempted to improve performance here by changing the network trained and tuning the hyperparemeters and running the training algorithm for more iterations.

Future work for the task of selecting ROIs in WSIs include adapting the model for other reward functions representing different tasks, such as object detection or finding patches rated as likely to be from an HCC (or CC) sample with above 90\% accuracy according to a classifier. We could also extend our candidate regions from tiles on a lattice to any rectangular region of the image by learning to output bounding boxes, similar to the Macias paper \cite{Maicas}. This could provide an augmented reality experience where pathologists can see in real-time which patches the model is selecting and get a sense of what it is looking for.

\section{Contributions \& Acknowledgements}
Rui primarily focused on building the baseline and the Deep-Q network training aspects of this paper. He also wrote some of the data preprocessing pipeline.

Jon primarily focused on the CNN + Beam search aspects of this paper. He also wrote some of the preprocessing pipeline.

Thanks to Amir Kiani and Dr. Shen for their assistance with the dataset and evaluation for our DQN. 

Outside sources: We used OpenAI's \hyperlink{https://github.com/openai/gym}{gym framework} to construct the environment for our Reinforcement Learning problem. We also used the deep cluster of the Stanford Machine Learning group to help train our models.

{\small
\bibliographystyle{ieee}

\bibliography{egbib}
}

\end{document}